\documentclass{article}

\usepackage{epsfig}
\newcommand{\beq}{\begin{equation}}
\newcommand{\eeq}{\end{equation}}
\newcommand{\beeq}{\begin{eqnarray}}
\newcommand{\eeeq}{\end{eqnarray}}

\begin{document}

\begin{center}\bf
HADRON VERTICES IN COMPOSITE SUPERSTRING MODEL
\bigskip
V.A.Kudryavtsev\\
Petersburg Nuclear Physics Institute \\

\bigskip

Abstract\\

 Hadron vertices for u,d,s, quark flavours are formulated in
terms of interacting composite strings. The vertices for emission of
$\pi, K $-mesons and nucleons are presented. \vspace{1pc}
\end{center}

\newpage
\section{String models for hadrons and composite strings. }

An essential interest in string description of hadron interactions
has  arised as far back as forty years ago (the Nambu string
\cite{1} and dual resonance models initiated by Veneziano's work
\cite{2}) due to the remarkable universal linearity of Regge
trajectories $\alpha(t)$ for meson and baryon resonances \cite{3}
and \cite{4}. $J=\alpha(M^2)=\alpha(0)+\alpha^\prime_H M^2$
$(\alpha^\prime_H\approx0.85\,$GeV$^{-2})$;  where J,M are spin and
mass of a resonance. Now we have these trajectories up to $J=5$ and
states for not only leading (n=0) but for second (n=1), for third
(n=2) and even for fourth (n=3)  daughter trajectories
$J_n=\alpha(0)-n+\alpha^\prime_HM^2$ n=0,1,2,3... . See \cite{4}.

  However attempts to build the string model for hadron interactions have
not been succesful since consistent i.e. compatible with unitarity
models for relativistic quantum strings \cite{5},\cite{6} have
required the intercept of leading meson trajectory $\alpha(0)$ to be
equal to one. A shift of this value from one has led to
contradiction  with unitarity since then states with negative norms
have appeared. The
 leading $\rho$-meson trajectory has the intercept to be equal
to one half approximately. Just this reason has led to superstring
models for nonhadrons:  for massless gluons (open strings) on the
trajectory $J=1+\alpha^\prime_PM^2$ and for massless gravitons
(closed strings) on the trajectory$J=2+1/2\alpha^\prime_PM^2$.

     A generalization of classical multireggeon (multistring)
   vertices \cite{7} has been suggested by author in 1993 \cite{8} as a new
   solution of duality equations for many-string vertices in \cite{7}.
   These string amplitudes have been used for description of
   interaction of many $\pi$-mesons \cite{8}.
   New string vertices \cite{9} give a new geometric picture for interactions
   of strings which has a natural description in terms of
   composite strings and three two-dimensional surfaces for moving
   open string instead of usual one.
     Additional edging two-dimensional surfaces carry quark
    quantum numbers (flavour, spin, chirality). This composite
    string  construction reminds two similar other composite objects:
    a gluon string with two pointlike quarks at ends of it
    or simplest case of a string ending at two
    branes when they are themselves some strings.
       It is not surprisingly as we shall see further that we have here
     a possibility for N=3 extended superconformal Virasoro symmetry
     for open composite strings. Let us note that we have no supersymmetry
     in the Minkovsky (target) space for this model. The topology of interacting
     composite strings allows to solve the problem of the intercept
     $\alpha(0)$ for leading meson trajectory and to shift  the
     value of this intercept to one half without breakdown of the
     extended superconformal Virasoro symmetry for composite strings
     due to non-vanishing conformal weights for fields on both edging two-dimensional surfaces.

\section{Formulation of composite string model. Simple vertices for interacting composite strings.}

   We repeat here and in the following section formulations of vertices and generators of symmetries for the critical
N=3 superconformal composite superstring model \cite{9} with some
corrections for formulation of supergauge generators $\widetilde
G_r$ and hence for the critical condition.

      For investigation of
composite superstrings it is more appropriate to move from
multi-string vertices to more simple vertices $\hat{V}_i$
corresponding to emission of ground states in an amplitude $A_N$ of
interaction of N strings. In this case of composite superstrings
operator vertices will contain additional ( to usual
$\partial{X_{\mu}}$ and $H_{\mu}$ fields on the basic
two-dimensional surface ) fields on edging surfaces :
  $Y_{\mu}$ and its superpartner  $f_{\mu}$ with Lorentz indices $\mu$ = 0,1,2,3 .
We include other edging fields ( J and its superpartner $\Phi$ )
which carry internal quantum numbers ( isospin and other flavour
currents) on edging surfaces and similar I, $\Theta$ - fields on the
basic two-dimensional surface.
    Since the edging fields are propagating only on the  own edging
 surfaces it is convenient to introduce vacuum states for the fields on the separate edging surfaces and
 to write $A_N$ in equivalent form with help of these vacuum states:
\begin{eqnarray}
&& A_N\!=\!\int\prod dz_i
\langle0^{(1,2)}|\hat{V}_{12}(z_1)\langle0^{(3)}|\hat{V}_{23}(z_2)\times
\nonumber \\
&&\langle0^{(4)}| \hat{V}_{34}(z_3)|0^{(2)}\rangle\
\langle0^{(5)}|... |0^{(i-1)}\rangle\ \hat{V}_{i,i+1}(z_i)\times \nonumber \\
&&\langle0^{(i+2)}|... |0^{(N-2)}\rangle\
\hat{V}_{N-1,N}(z_{N-1})|0^{(N-1)}\rangle\ \times \nonumber  \\
&&\hat{V}_{N,1}(z_N)|0^{(N,1)}\rangle\
\end{eqnarray}
  This form(1) excludes this amplitude from the set of additive string models of the Lovelace's paper \cite{5}
and leads  to the topology of composite string models
\cite{8},\cite{9}. Now we are ready to formulate the vertex operator
$\hat{V}_{i,i+1}(z_i)$ (Figure 1.) for this composite string model:
\begin{eqnarray}
&& \hat{V}_{i,i+1}(z_i)\ =\ z_i^{-L_0}\left[G_r,\hat W_{i,i+1}
\right]z_i^{L_0}\ ,\nonumber\\
&&\hat W_{i,i+1}=\hat{R}^{out}_{i}\hat{R}_{NS}\hat{R}^{in}_{i+1}
\end{eqnarray}

\begin{figure}[htb]
\begin{center}
\epsfig{figure=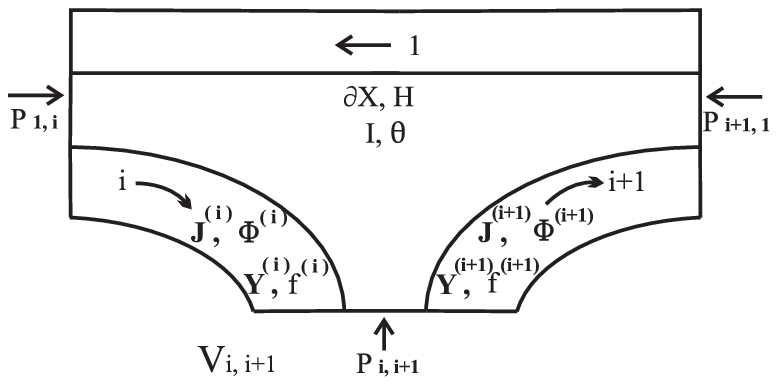,width=0.50\textwidth,clip=}
\caption{\label{fig1}}
\end{center}
\end{figure}

 The operators $\hat{R}^{out}_{i}$and $\hat{R}^{in}_{i+1}$
are defined by fields on i-th and (i+1)-th  edging surfaces. The
operator $\hat{R}_{NS}$ is defined by fields on the basic surface.
They have the same structure as the operator $:\exp{ip_i X(1)}:$ of
classical string models  for both $Y$ and $J$ -fields:
\begin{eqnarray}
&& \hat{R}^{out}_{i}= \exp{(\xi_{i}\sum_n \frac {J^{(i)}_{-n}}{n})}\
\exp{(k_{i} \sum_n \frac {Y^{(i)}_{-n}}{n})}\ \times \nonumber \\
&& \exp{(ik_{i} Y^{(i)}_0)} \widetilde\lambda^{(+)}_i \exp{(-k_{i}
\sum_n \frac {Y^{(i)}_{n}}{n})}\ \times \nonumber \\
&& \exp{(-\xi_{i} \sum_n \frac {J^{(i)}_{n}}{n})}\;
\end{eqnarray}
And we have the similar expression for   $\hat{R}^{in}_{i+1}$:
 \begin{eqnarray}
&&\hat{R}^{in}_{i+1}= \exp{(-\xi_{i+1}\sum_n \frac
{J^{(i+1)}_{-n}}{n})}\ \exp{(-k_{i+1} \sum_n \frac
{Y^{(i+1)}_{-n}}{n})}\ \times \nonumber \\
&&\exp{(-ik_{i+1} Y^{(i+1)}_0)}
\lambda^{(-)}_{i+1} \exp{(k_{i+1} \sum_n \frac {Y^{(i+1)}_{n}}{n})}\ \times \nonumber \\
&&\exp{(\xi_{i+1} \sum_n \frac {J^{(i+1)}_{n}}{n})}\
\end{eqnarray}
\begin{eqnarray}
 \sum_i k_i=0
\end{eqnarray}
\begin{eqnarray}
&&\hat{R}^{(NS)}_{i,i+1}=  \exp{(-\zeta_{i,i+1}\sum_n \frac
{I_{-n}}{n})}\ \exp{(-p_{i,i+1}\sum_n \frac {a_{-n}}{n})}\  \times \nonumber \\
&&\exp{(-ip_{i,i+1}X_0)}\Gamma_{i,i+1} \exp{(p_{i,i+1}\sum_n \frac
{a_{n}}{n})}\  \times \nonumber \\
&&\exp{(\zeta_{i,i+1}\sum_n \frac {I_{n}}{n})}\
\end{eqnarray}
   Here we have introduced $\lambda_{\alpha}$ operators to be carrying
quark flavours and quark spin degrees of freedom.
\begin{eqnarray}
\langle0|\widetilde\lambda^{(+)}=0; \lambda^{(-)}|0\rangle=0
\end{eqnarray}
\begin{eqnarray*}
&&
\left\{\tilde\lambda_{\alpha}^{(-)},\lambda_{\beta}^{(+)}\right\}\
=\  \delta_{\alpha,\beta}\ ; \quad \tilde\lambda\ =\ \lambda T_0 ,\\
&& T_0=\gamma_0\otimes\tau_2\ ;
\end{eqnarray*}
\begin{eqnarray}
&&\hat{p}_{i,i+1}=\hat{\beta}^{(i+1)}_{in} \hat{p}_{i+1} +
\hat{\beta}^{(i)}_{out} \hat{p}_{i}= \nonumber \\
&& \hat{\beta}^{(i+1)}_{in} k_{i+1} - \hat{\beta}^{(i)}_{out} k_{i}
\end{eqnarray}
\begin{eqnarray}
\hat{\zeta}_{i,i+1}=
\hat{\alpha}^{(i+1)}_{in}\hat{\xi}_{i+1}+\hat{\alpha}^{(i)}_{out}\hat{\xi}_{i}
\nonumber \\
\hat{\xi}_{i}= \widetilde\lambda^{(+)}_{i}\xi \lambda^{(-)}_{i};\
\end{eqnarray}
       Here $\xi$ is some universal matrix over quark flavours.
  So we give some relation between of momenta (charges) which flow
into the basic surface and into edging surfaces.  Namely operators
$\hat{\beta}(\hat{\alpha})$ define  fractions of i-th and (i+1)-th
momenta (charges) for the basic surface:
\begin{eqnarray}
\hat{\beta}^{(i)}_{out}=\widetilde\lambda^{(+)}_{i} \beta_{out}\lambda^{(-)}_{i};\nonumber\\
\hat{\alpha}^{(i)}_{out}=\widetilde\lambda^{(+)}_{i}\alpha_{out}\lambda^{(-)}_{i};
\end{eqnarray}
\begin{eqnarray}
[\beta,\alpha ]=0; \beta ^2 = \alpha^2=1
\end{eqnarray}

\section{ Extended Virasoro superconformal symmetries, supercurrent conditions and critical case for composite superstrings}

Main symmetry of any string model is the superconformal
 symmetry to be defined by the Virasoro operators $G_r$.
    For composite superstring model we consider the set of states
and of superconformal generators for the i-th section between the
$\hat{V}_{i-1,i}$ vertex and $\hat{V}_{i,i+1}$ vertex in (1)(see
Figure 2.).

\begin{figure}[htb]
\begin{center}
\epsfig{figure=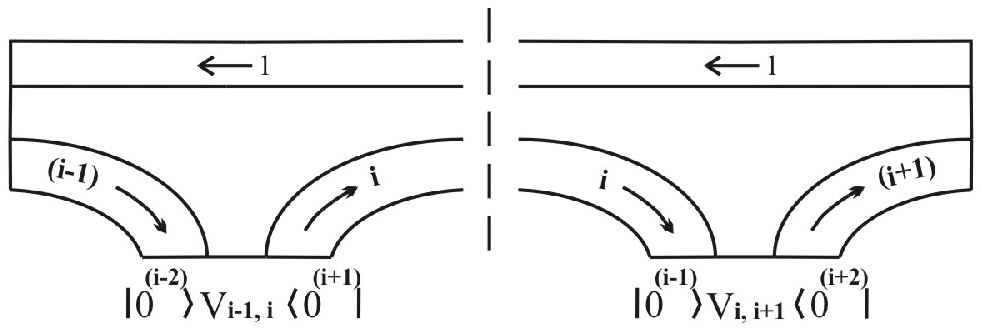,width=0.50\textwidth,clip=}
\caption{\label{fig2}}
\end{center}
\end{figure}

        Namely we have fields on (i-1), i, (i+1) edging surfaces :\\
        $(Y^{(i-1)},f^{(i-1)});(J^{(i-1)},\Phi^{(i-1)})$;$(Y^{(i)},f^{(i)});(J^{(i)}$ \\
        $\Phi^{(i)})$; $(Y^{(i+1)},f^{(i+1)});(J^{(i+1)},\Phi^{(i+1)})$
        fields

in addition to  fields which are on the basic surface:\\
        $(\partial{X},H)$;$(I,\Theta)$;

   In order to consider our spectrum of states in more symmetric way
 regards left and right sides we introduce a set of auxiliary
 fields $(Y^{(a)},f^{(a)});(J^{(a)},\Phi^{(a)})$ instead of (i-1)-th and
 (i+1)-th fields in decomposition of 1 in the i-th section:
\begin{eqnarray}
 1=\sum |State(i-1,i,i+1)\rangle\ \langle State(i-1,i,i+1)|
\end{eqnarray}
Taking into account the operator $\langle 0^{(i+1)}|$ from the left
side and operator $|0^{(i-1)}\rangle\ $ from the right side we can
replace  (12) by the following sum:
\begin{eqnarray}
&&\sum |State (i-1,i)\rangle\ \langle State (i,i+1)|=  \nonumber \\
&&\delta (fields(i-1)- fields (a))\sum |State (a,i)\rangle\ \times  \nonumber \\
&&\langle State (i,a)| \times\delta (fields(a)-fields(i+1))
\end{eqnarray}
  Here we mean  for $\delta (fields(i-1)-fields (a))$
  (in the case of Y-fields):
\begin{eqnarray}
&&\delta (Y(i-1)- Y(a))=    \\
&&\sum_{[\lambda_1,\lambda_2,...]}\prod_{1}^{n}{\frac{(Y^{(i-1)}_{-n})^{\lambda_n}}
{\sqrt{\lambda_n !}}}|0^{(i-1)}\rangle\ \langle
0^{(a)}|\prod_{1}^{n}{\frac{(Y^{(a)}_n)^{\lambda_n}}{\sqrt{\lambda_n!}}}
\nonumber
\end{eqnarray}
and  for $\delta (fields(a)- fields (i+1))$ correspondingly :
\begin{eqnarray}
&&\delta (Y(a)- Y(i+1))=   \\
&&\sum_{[\lambda_1,\lambda_2,...]}
\prod_{1}^{n}{\frac{(Y^{(a)}_{-n})^{\lambda_n}}{\sqrt{\lambda_n
!}}}|0^{(a)}\rangle\ \langle
0^{(i+1)}|\prod_{1}^{n}{\frac{(Y^{(i+1)}_n)^{\lambda_n}}{\sqrt{\lambda_n!}}}
\nonumber
\end{eqnarray}

   It is worth to note that the sets of states on the left and on the right of the
vertex  $\hat{V}_{i,i+1}$ will be the same ones that allows to
connect corresponding decompositions unambiguously and to see
independence on the number of the section for this consideration.

    We shall consider this operator $\sum |State (a,i)\rangle\
   \langle State(i,a)|$ which represents 1 in the Fock space for
   $|State (a,i)\rangle\ $ and we shall extract spurious states in order to
find the physical states spectrum.

  Now superconformal generators $G_r$  can be defined as the
following ones:
\begin{eqnarray}
&& G_r=G^{Lor}_r + G^{Int}_r
\end{eqnarray}
\begin{eqnarray}
&&G^{Lor}_r=\frac{1}{2\pi}\int\limits_0^{2\pi}d\tau
[(H^{\mu}\frac {d}{d\tau} X_\mu + \hat{P}_\nu H^\nu)+  \nonumber \\
&&(Y^{(a)}_{\mu}f^{(a)\mu } + \hat{p}_{1} f^{(1)}+Y^{(i)}_\mu
f^{(i)\mu})] e^{-ir\tau}
\end{eqnarray}
\begin{eqnarray}
&& G^{Int}_r\!=\!\frac1{2\pi}\int\limits_0^{2\pi}d\tau [(I\Theta+
\xi_1\Phi^{(1)})+  \nonumber \\
&&(J^{(a)}\Phi^{(a)}+ J^{(i)}\Phi^{(i)})] e^{-ir\tau}
\end{eqnarray}
 Here we have a=i-1 (on the left side) or a=i+1 (on the right
    side).
      But unlike the Neveu-Schwarz model this composite string model
      has a new superconformal symmetry which defines by the
      following generators $\widetilde G_r$:
\begin{eqnarray}
&&\widetilde G_r=\widetilde G^{Lor}_r + \widetilde G^{Int}_r
\end{eqnarray}
\begin{eqnarray}
&&\widetilde G^{Lor}_r=  \\
&&\frac1{2\pi}\int\limits_0^{2\pi}d\tau [\partial{X}_\mu
\hat{\beta}^{(a)}f^{(a)\mu}+\partial{X}_\mu
\hat{\beta}^{(i)} f^{(i)\mu}+  \nonumber \\
&& \hat {p}_1 f^{(1)}+Y^{(a)}_\mu \hat{\beta}^{(a)}H^{\mu}+
Y^{(i)}_\mu
\hat{\beta}^{(i)}H^{\mu}- \nonumber \\
&&(Y^{(a)}_\mu\hat{\beta}^{(a)}\hat{\beta}^{(i)}f^{(i)\mu}+Y^{(i)}_\mu\hat{\beta}^{(i)}\hat{\beta}^{(a)}f^{(a)\mu})]e^{-ir\tau}\nonumber
\end{eqnarray}
\begin{eqnarray}
&&\widetilde G^{Int}_r=\frac1{2\pi}\int\limits_0^{2\pi}d\tau
[(I\hat{\alpha}^{(a)}\Phi^{(a)}+I\hat{\alpha}^{(i)}\Phi^{(i)})+  \nonumber \\
&&\xi_1\Phi^{(1)}+J^{(a)}\hat{\alpha}^{(a)}\Theta +
J^{(i)}\hat{\alpha}^{(i)}\Theta
-  \\
&&(J^{(a)}\hat{\alpha}^{(a)}\hat{\alpha}^{(i)}\Phi^{(i)}+J^{(i)}\hat{\alpha}^{(i)}\hat{\alpha}^{(a)}\Phi^{(a)}
)]e^{-ir\tau}  \nonumber
\end{eqnarray}
      The operators $\widetilde G_r$ give us the same commutation
   relations to the operator vertex $\hat{V}_{i,i+1}$ as $G_r$ ( here we have $a=i+1$ in $G_r$ and $\widetilde G_r$ ) :
\begin{eqnarray}
&&[\widetilde G_r,\hat{V}_{i,i+1}(1)] = [G_r,\hat{V}_{i,i+1}(1)];
\nonumber\\ &&[\widetilde G_r,\hat{W}_{i,i+1}] =
[G_r,\hat{W}_{i,i+1}]
\end{eqnarray}

Just here we have used the definite relations (8)-(11) for momenta
and charges in the operator vertex $\hat{V}_{i,i+1}$.
  Taking into account the expressions (16)-(21) we
can derive the corresponding commutation relations for $\widetilde
G_r$ and $G_r$:

\begin{eqnarray}
\{G_r,G_s\}\ =\ 2L_{r+s}+\frac D2\left(r^2-\frac14\right)\delta_{r,-s} \\
D=3d^{Lor}+3d^{Int}\\
 \{G_r,\widetilde G_s\}\ = 2\widetilde L_{r+s};
\end{eqnarray}
here $d^{Lor}$ is the number of $Y$(or $X$)-components,\\
$d^{Int}$ is the number of  $J$(or $I$)-components.
\begin{eqnarray}
\{\widetilde G_r,\widetilde G_s\}\ = 4L_{r+s}-2\widetilde L_{r+s}
 +\frac {\widetilde D}2 \left(r^2-\frac14\right)\delta_{r,-s}\\
\widetilde D=6d^{Lor}+6d^{Int}
\end{eqnarray}
\begin{eqnarray}
[L_n,L_m]=(n-m)L_{n+m}+\frac{D}{8}n(n^2-1)\delta_{n,-m}
\end{eqnarray}
\begin{eqnarray}
[L_n,\widetilde L_m]=(n-m)\widetilde L_{n+m}
\end{eqnarray}
\begin{eqnarray}
&&[\widetilde L_n ,\widetilde L_m]=2(n-m)L_{n+m}- (n-m) \widetilde
L_{n+m} + \nonumber \\
&&\frac{\widetilde D}{8}n(n^2-1)\delta_{n,-m}
\end{eqnarray}
\begin{eqnarray}
[L_n,G_r] =\left( \frac n2-r \right)G_{n+r}
\end{eqnarray}
\begin{eqnarray}
[L_n,\widetilde G_r] =\left( \frac n2-r \right)\widetilde G_{n+r}
\end{eqnarray}
\begin{eqnarray}
[\widetilde L_n ,\widetilde G_r] =\left(n-2r \right)G_{n+r}-\left(
\frac n2-r \right)\widetilde G_{n+r}
\end{eqnarray}

  Due to this algebra and equations (22) we are able to prove
as earlier in classical models that both $G_r$ and $\widetilde G_r$
operators generate spurious
  states.
This commutation agebra allows to extract the independent
combinations of operators $G_r$ and $\widetilde G_r$ which define
the spectrum of spurious states.
  It is possible to extract the independent
combinations of operators $G_r$ and $\widetilde G_r$ which define
the spectrum of spurious states:
\begin{eqnarray}
&&G^I_r= \frac 13(\widetilde G_r+2G_r);G^{II}_r=(G^{Lor}_r -
\widetilde G^{Lor}_r);\nonumber\\ &&G^{III}_r=(G^{Int}_r -
\widetilde G^{Int}_r)
\end{eqnarray}

     Let us notice that our construction of vertices
(2),(3)-(6) according to (8),(9) contains some definite combinations
of fields with Lorentz indices:
\begin{eqnarray}
k_i\widetilde f^{(i)}= k_i (f^{(i)} + \hat{\beta}_{i} H);\nonumber\\
k_i \widetilde Y^{(i)}= k_i (Y^{(i)}+ \hat{\beta}_i \partial X )
\end{eqnarray}
and with internal quantum numbers:
\begin{eqnarray}
\zeta_i \widetilde \Phi^{(i)}=
\zeta_i(\Phi^{(i)}+\hat{\alpha}_i\Theta);\nonumber\\
  \zeta_i\widetilde J^{(i)}= \zeta_i(J^{(i)}+\hat {\alpha}_i I)
\end{eqnarray}

  Let us notice that all combinations (35) and (36) commute
  with operators $G^{II}_r; G^{III}_r$.

      Let us consider the construction of the spectrum generating
algebra for this composite superstring by similar way as in
classical string models [6]. For the given i- th section (
betweeen $V_{i-1,i}$ and $V_{i,i+1}$) we have fields on i,i-1,i+1
edging surfaces and fields on the basic surface.
    Spurious states for this basis are defined by products of
  operators $G^I_r,G^{II}_r,G^{III}_r$ and
  $L^I_n,L^{II}_n,L^{III}_n$. But only these states are not able to save
  from negative norms the spectrum of physical states as it has
  taken place for usual classical string models since the capacity
 of those of them which have negative norms is not enough.
For the Fock space under
 consideration we can obtain states with negative norms not only as the
 powers of time components of the $\partial X $
and $H $ fields on the basic surface but as odd powers of other
time-like components :$k_{a}\widetilde f^{(a)},k_{i}\widetilde
f^{(i)}$ and $k_{a} \widetilde Y^{(a)},k_i \widetilde Y^{(i)}$.

     Additional conditions for the composite string model are
 required in order to eliminate all negative norms from the spectrum
 of physical states. There is a simple solution for it. We shall
 require as gauge conditions the supercurrent
  conditions generated  by   $k_i \widetilde {f}^{(i)}$.

  Namely we shall take the following constraints for our vertices:
\begin{eqnarray}
[k_i \widetilde Y^{(i)}_n,\hat{W}_{i,i+1}] =
[\hat{W}_{i,i+1},k_{i+1} \widetilde Y^{(i+1)}_n]=0
\end{eqnarray}
  Then we shall have enough states of negative norms generated by
all gauge constraints. The equations (37) lead  to the conditions:
\begin{eqnarray}
k_i^2 \rightarrow 0;k_{i+1}^2 \rightarrow 0;(k_ik_{i+1})\rightarrow
0;
\end{eqnarray}

   So our gauge supercurrents are independent and nilpotent ones:
\begin{eqnarray}
[k_i \widetilde Y^{(i)}_n,k_i \widetilde Y^{(i)}_m] =0;
 [k_{i+1} \widetilde Y^{(i+1)}_n,k_i \widetilde Y^{(i)}_m]=0
\end{eqnarray}

    Let us notice that our choice for additional gauge conditions
is appropriate for emission of $\pi$-mesons  (the case of usual
quarks). It gives an explanation for massless $\pi$-mesons and
correct amplitudes for $\pi$-mesons interaction [9]. But other quark
flavours  bring us to gauge supercurrent constrains which contain
not only fields with Lorentz indices $\widetilde Y^{(i)}$ but and
some part of  fields $\widetilde J^{(i)}$  for internal quantum
numbers. This part is vanishing for the case of usual $\pi$ -
mesons.

   Now we are able to build spectrum generating
algebra (SGA) for our set of states in the same manner as for the
Neveu- Schwarz string model in [6] with help of the operators of
type of vertex operators (2) of the conformal spin j to be equal to
one.

      We shall use the light-like vectors  $k^{(LI)}_i$  from our vertices
  ($(k^{(LI)}_i)^2=0$) and consider a state of the
 generalized momentum  $P^{(gen)}=p_0 + Nk^{(LI)}_i$.

\begin{eqnarray}
&&\frac {p_0^2}2= -1;(k^{(LI)}_i)^2=0;(k^{(LI)}_i p_0)=1 \nonumber \\
&&(k^{(LI)}_{i-1})^2=0;(k^{(LI)}_ {i+1}   )^2=0;\\
&&(k^{(LI)}_ik^{(LI)}_{i+1})\rightarrow 0;
(k^{(LI)}_ik^{(LI)}_{i-1})\rightarrow 0;\nonumber
\end{eqnarray}
    Transversal components of $k^{(LI)}_i,p_0$ are vanishing $(p_0)_a
=(k^{(LI)}_i)_a=0$. The generalized mass of this state is given by :
\begin{eqnarray}
\frac {M^{(gen)2}}2=\frac{(p_0+ Nk_i^{(LI)})^2}2 =-1+N
\end{eqnarray}

We define the transversal operators of SGA as corresponding vertex
operators.

         All transversal SGA operators satisfy simple commutation algebra:
\begin{eqnarray}
&&[(S^{(i)}_a)_n,(S^{(j)}_b)_m]=m\delta^{i,j}\delta_{a,b}\delta_{m+n,0} \nonumber  \\
&&[(S^{(i)}_a)_n,(B^{(j)}_b)_r]=0; \nonumber \\
&&\{(B^{(i)}_a)_r,(B^{(j)}_b)_s\}=\delta^{i,j}\delta_{a,b}\delta_{m+n,0}\nonumber
\end{eqnarray}
 So we can construct similarly to the DDF states transversal
 states $|Phys\rangle$ from powers of the transversal SGA operators:
\begin{eqnarray}
|Phys\rangle =
 \prod {((S^{(i)}_a)_{-n})^{\lambda (a,n)}}...|\Psi_0\rangle
\end{eqnarray}
These states  satisfy all necessary gauge conditions.

    Let us notice that all transversal SGA operators on the left side with
    (i-1)- and (i)- operators
( (i+1)-operators are vanishing there)  can be defined
 with replacement of
 all (i)-fields to (i-1)-fields and vice versa of all (i-1)-fields to (i)-fields.
  It is true and for
all transversal SGA operators on the right side with (i+1)- and (i)-
operators  ( (i-1)- operators are vanishing there ). They can be
defined with replacement of all (i)-fields to (i+1)-fields and vice
versa of all (i+1)-fields to (i)-fields. This possibility to
reformulate these sets of states allows to move from states of i-th
section under consideration to states in (i-1)-th section and so on.
Hence our consideration can be carried out up to ends of our
amplitude (1) and does not depend on the number i of this section.

   Moving from these DDF type states to arbitrary states  we can
obtain them as usually with help of ordered powers of the conformal
generators $G^I_r,L^I_n;_n;G^{II}_r,L^{II}_n;G^{III}_r,L^{III}_n$
and  of  powers  of  the  supercurrent operators \\$k^{(LI)}_i
\widetilde Y^{(LI)(i)}$, $k^{(LI)}_i \widetilde f^{(LI)(i)}$ acting
on $|Phys\rangle$  states:

\begin{eqnarray}
&&(G^I_{\frac {-1}2})^{\lambda (1)}(G^I_{\frac {-3}2})^{\lambda
(3)}...(L^I_{-1})^{\mu (1)}(L^I_{-2})^{\mu (2)}...\nonumber \\
&&\prod_r (k^{(LI)}_{i-1} \widetilde f^{(LI)(i-1)}_{-r})^{\gamma
(i-1,r)} \nonumber \\
&&\prod_n (k^{(LI)}_{i-1} \widetilde Y^{(LI)(i-1)}_{-n})^{\delta
(i-1,n)} \nonumber \\
&&\prod_r (k^{(LI)}_{i} \widetilde f^{(LI)(i)}_{-r})^{\gamma (i,r)}
\nonumber \\
&&\prod_n (k^{(LI)}_{i} \widetilde Y^{(LI)(i)}_{-n})^{\delta
(i,n)}|Phys\rangle
\end{eqnarray}

     Then we can repeat considerations in the Neveu-Schwarz
model \cite{6}] for  the  theorem about absence of ghosts in the
spectrum of physical states in our case for the critical value of
the number of effective dimensions and taking into account the
conditions (40).

In critical case the operators $G^I_{\frac {1}2}$ and
\begin{eqnarray}
G^I_{\frac{3}2}+ 2 (G^I_{\frac {1}2})^3=\nonumber\\
=G^I_{\frac{3}2}+ \frac 2{3} (\widetilde G_{\frac {1}2}+ 2G_{\frac
{1}2})( \widetilde L_1 + 2L_1 )= G'^I_{\frac{3}2}
\end{eqnarray}
define null states:
\begin{eqnarray} \{G'^I_{\frac{3}2},G^I_{\frac{-1}2}\}=0
\end{eqnarray}
\begin{eqnarray}
|S_{\frac{-3}2}\rangle = G'^I_{\frac{-3}2}|Phys\rangle;\langle S_{\frac{3}2}|S_{\frac{-1}2}\rangle =0 \\
| S_{\frac{-1}2}\rangle =
G^I_{\frac{-1}2}|Phys\rangle;\langle|S_{\frac{1}2}
|S_{\frac{-1}2}\rangle =0\nonumber
\end{eqnarray}

\begin{eqnarray}
\langle|S_{\frac{3}2} |S_{\frac{-3}2}\rangle =0
\end{eqnarray}
    The critical case corresponds to the condition (46).
    It requires
 the condition (47) to be satisfied:
\begin{eqnarray}
L_0=\widetilde L_0=\frac{1}2
\end{eqnarray}

   and  definite values of numbers of fields:
\begin{eqnarray}
d_{crit}= 2d^{Lor} + 2d^{Int} =10
\end{eqnarray}

    For the critical case we can prove by the same way as in the
Neveu-Schwarz model that the norms of all physical states are
nonnegative if all constraints for physical states are fulfilled.
       That means
four  fields for all Y-fields i.e. $Y^{(i)}_{\mu},\mu=0,1,2,3$ and
 one component for  $J^{(i)}$-fields.

\section{Hadron vertices for u,d,s quark flavours in tree hadron amplitudes}

   Let us consider simple composite critical superstring vertices for
usual u,d and s quark flavours in arbitrary tree amplitudes.

    Our choice for $\pi$-meson emission vertex coincides with the
simplest vertex (2)-(6)

 with $k_{i}^2=k_{i+1}^2=\mu^2$;$k_{i}k_{i+1}=0 $ and $\mu ^2\rightarrow 0$ .
  The last conditions provide the supercurrent conditions (38) and leads to $m_{\pi}^2=0$.

    We can propose some simple choice for $\beta_{in},\beta_{out}$:
\begin{eqnarray}
\beta_{in}^{(i+1)}=a\frac{\hat{k}_{i+1}}{\mu} + b\gamma_5;
\beta_{out}^{(i)}=a\frac{\hat{k}_{i}}{\mu}-b\gamma_5;\nonumber\\
 a=\cos{\phi};b=\sin{\phi};
\end{eqnarray}
  Let us notice that value of $\phi$  defines the fraction of momentum to be
flowing  into two-dimensional surface for the closed string sector
and therefore $\phi^2 \sim 10^{-38}$.
      As it is proposed  $\lambda^{(-)}_{i+1}$ and  $\lambda^{(+)}_{i}$
are eigenfunctions of operators $\hat{\beta_{in}^{(i+1)}}$  and
$\hat{\beta_{out}^{(i)}}$ correspondingly :

\begin{eqnarray}
\hat{\beta_{out}^{(i)}}\lambda^{(+)}_{(\eta_{i})}|0\rangle=
\eta_{i}\lambda^{(+)}_{(\eta_{i})}|0\rangle ; \eta_{i}=\pm 1   \\
 \langle0|\widetilde \lambda^{(-)}_{(\eta_{i+1})}\hat{\beta_{in}^{(i+1)}}=
\eta_{i+1}\langle0|\widetilde \lambda^{(-)}_{(\eta_{i+1})};
\eta_{i+1}=\pm 1
\end{eqnarray}

  So we have the following $\pi$- meson emission vertex:
\begin{eqnarray}
&& \hat{V}_{i,i+1}(z_i)\ =\ z_i^{-L_0}\left[G_r,\hat W_{i,i+1}
\right]z_i^{L_0}\ ,\nonumber\\
&&\hat W_{i,i+1}=\hat{R}^{out}_{i}\hat{R}_{NS}\hat{R}^{in}_{i+1}
\end{eqnarray}
\begin{eqnarray}
&& \hat{R}^{out}_{i}= \nonumber\\
&&\exp{(\xi_{i}\sum_n \frac {J^{(i)}_{-n}}{n})}\
\exp{(k_{i} \sum_n \frac {Y^{(i)}_{-n}}{n})}\ \times \nonumber \\
&& \exp{(ik_{i} Y^{(i)}_0)} \widetilde\lambda^{(+)}_i \exp{(-k_{i}
\sum_n \frac {Y^{(i)}_{n}}{n})}\ \times \nonumber \\
&& \exp{(-\xi_{i} \sum_n \frac {J^{(i)}_{n}}{n})}\;
\end{eqnarray}
\begin{eqnarray}
&&\hat{R}^{in}_{i+1}=\nonumber\\
&& \exp{(-\xi_{i+1}\sum_n \frac {J^{(i+1)}_{-n}}{n})}\
\exp{(-k_{i+1} \sum_n \frac
{Y^{(i+1)}_{-n}}{n})}\ \times \nonumber \\
&&\exp{(-ik_{i+1} Y^{(i+1)}_0)}
\lambda^{(-)}_{i+1} \exp{(k_{i+1} \sum_n \frac {Y^{(i+1)}_{n}}{n})}\ \times \nonumber \\
&&\exp{(\xi_{i+1} \sum_n \frac {J^{(i+1)}_{n}}{n})}\
\end{eqnarray}
\begin{eqnarray}
&&\hat{R}^{(NS)}_{i,i+1}=\nonumber\\
&&  \exp{(-\zeta_{i,i+1}\sum_n \frac
{I_{-n}}{n})}\ \exp{(-p_{i,i+1}\sum_n \frac {a_{-n}}{n})}\  \times \nonumber \\
&&\exp{(-ip_{i,i+1}X_0)}\Gamma_{i,i+1} \exp{(p_{i,i+1}\sum_n \frac
{a_{n}}{n})}\  \times \nonumber \\
&&\exp{(\zeta_{i,i+1}\sum_n \frac {I_{n}}{n})}\
\end{eqnarray}
   We require $\xi_{i}^2=\xi_{i+1}^2=\frac {1}2$ in order to have
the conformal spin of this vertex to be equal to one.

    The product
 $\widetilde\lambda^{(+)}_{(\eta_{i})}...\Gamma_{i,i+1}...\lambda^{(-)}_{(\eta_{i+1})}$
can be presented for the $\pi$- meson emission vertex in the
following way:
\begin{eqnarray}
\widetilde\lambda^{(+)}_{\beta,(\eta_{i})}...\Gamma_{i,i+1}...
\lambda^{(-)}_{\alpha,(\eta_{i+1})}=\nonumber\\
\sum_{\eta_{i},\eta_{i+1},\alpha,\beta}\widetilde\lambda^{(+)}_{\beta,(\eta_{i})}...\epsilon_{\eta_{i},\eta_{i+1}}
\tau^{\pi(ij)}_{\beta \alpha}... \lambda^{(-)}_{\alpha,(\eta_{i+1})}
\end{eqnarray}
  Here $\beta;\alpha =1,2$ are isotopic indices and $\tau^{\pi(ij)}$
  is a corresponding isotopic Pauli matrix.

  For $A_4$ we have from (1)
\begin{eqnarray}
&& A_4\!=\!\int\prod dz_i
\langle0^{(1,2)}|\hat{V}_{12}(z_1)\langle0^{(3)}|\hat{V}_{23}(z_2)\times
\nonumber \\
&&\langle0^{(4)}| |0^{(2)}\rangle\
\hat{V}_{3,4}(z_{3})|0^{(3)}\rangle\
\hat{V}_{4,1}(z_4)|0^{(4,1)}\rangle\
\end{eqnarray}

     For $\pi+\pi \rightarrow \pi+\pi $ we have
\begin{eqnarray}
&&A_4 = g^2\int_0^1 dx x^{-\frac3{2}}x^{(-\frac{p_{13}^2 }2+
\frac{\xi_{1}^2}2+\frac{\xi_{3}^2}2-\frac{k_{1}^2}2-\frac{k_{3}^2}2+\frac{\zeta_{13}^2}2)+\frac1{2}} \nonumber\\
&&(1-x)^{-p_{23}p_{34}+\zeta_{23}\zeta_{34}+k_{3}^2-\xi_{3}^2-1} \\
&&(-p_{23}p_{34}+\zeta_{23}\zeta_{34}+k_{3}^2-\xi_{3}^2)
Tr(\Gamma_{12}\Gamma_{23}\Gamma_{34}\Gamma_{41}) \nonumber
\end{eqnarray}
So we obtain this amplitude as a simple beta function
\begin{eqnarray}
&&A_4 = g^2 \frac{\Gamma (1- \alpha^t_0-\frac1{2}t)\Gamma (1-
\alpha^s_0-\frac1{2}s)} {\Gamma
(1-\alpha^t_0-\frac1{2}t-\alpha^s_0-\frac1{2}s))}
\nonumber \\
&&Tr(\Gamma_{12}\Gamma_{23}\Gamma_{34}\Gamma_{41})
\end{eqnarray}
 with $t=p_{13}^2, s=p_{34}^2;$\\
 $\alpha^t_0=1-\frac{\xi_{1}^2}2-\frac{\xi_{3}^2}2+\frac{k_{1}^2}2+\frac{k_{3}^2}2-\frac{\zeta_{13}^2}2);$\\
$\alpha^s_0=-\frac{p_{23}^2}2+\frac
{p_{34}^2}2-\zeta_{23}\zeta_{34}-k_{3}^2+\xi_{3}^2$;\nonumber\\

$k_{i}^2=0$;$\xi_{i}^2=\xi^2=\frac1{2}$  and hence
$\zeta_{13}=\xi_{1}+\xi_{3}=0,\zeta_{23}=\zeta_{34}=0$;
$p_{23}^2=p_{34}^2=m_{\pi}^2=0$ \\ and
$\alpha^t_0=\alpha^s_0=\frac1{2}$

       After this natural choice for the $\pi$- meson emission
 vertex we can not build the K- meson emission vertex similarly
 without a lost of the supercurrent conditions (37) for the s-quark edging
 surface and hence with the breakdown of our construction of the spectrum generating algebra
 and then with appearance of states of negative norms in the physical
 spectrum.\\

      But it is possible to move to another form for
 $\hat W_{i,i+1}$ in the case of K-mesons without a lost of the supercurrent
 conditions:
\begin{eqnarray}
&& \hat{V}_{i,i+1}(z_i)\ =\ z_i^{-L_0}\{G_r, \widetilde
W_{i,i+1}\}z_i^{L_0}\ ,\nonumber\\
&&\widetilde W_{i,i+1}=\left[G_r,\hat W_{i,i+1}\right]\\
&& \hat
W_{i,i+1}=\hat{R}^{out}_{i}\hat{R}_{NS}\hat{R}^{in}_{i+1}\nonumber
\end{eqnarray}
   Here we have i-th edging surface for usual (u,d) quark flavours
with $k_{i}^2=\mu^2$;$k_{i}k_{i+1}=0;\mu ^2\rightarrow
0$;$\xi_{i}^2=\frac1{2}$ and i+1-th edging surface for s-quark
flavour.\\

    There are two orthogonal light-like supercurrent
conditions:
 the old one
\begin{eqnarray}
k_i\widetilde f^{(i)}= k_i (f^{(i)} + \hat{\beta}_{i} H);\nonumber\\
k_i \widetilde Y^{(i)}= k_i (Y^{(i)}+ \hat{\beta}_i \partial X )
\end{eqnarray}
and the second one
\begin{eqnarray}
&&\widetilde f_s = k_{i+1}(f^{(i+1)} + \hat{\beta}_{i+1}H)
-\xi_i \Phi^{(i)}\nonumber\\&&+ \xi_{i+1}\Phi^{(i+1)}+(\xi_{i+1}\hat{\alpha}_{i+1}-\xi_i\hat{\alpha}_i)\Theta;\nonumber\\
&& \widetilde Y_s =- k_{i+1}(Y^{(i+1)}+ \hat{\beta}_{i+1}\partial X )- \xi_i J^{(i)}\nonumber\\
&&+\xi_{i+1} J^{(i+1)} + (\xi_{i+1}\hat
{\alpha}_{i+1}-\xi_i\hat{\alpha}_i)I
\end{eqnarray}

   We require
\begin{eqnarray}
-k_{i+1}^2+\xi_{i}^2+\xi_{i+1}^2-\xi_{i}\xi_{i+1}=0
\end{eqnarray}
   and $k_{i}^2=0$
   in order to have
the conformal spin of this vertex to be equal to one and the
light-likeness of (62),(63) simultaneously.

   For $\xi_{i}^2=\frac 1{2}$ and for the minimal value of $\frac {k_{i+1}^2}2=\frac {m_K^2}{2}$ we have
\begin{eqnarray}
k_{i+1}^2=k_s^2=\frac 3{8};\xi_{i+1}=\xi_s=\frac 1{2}\xi_{i}
\end{eqnarray}
It corresponds to $m_K\approx 474Mev$.

    We take here in the K-meson emission vertex as for $\pi$-mesons
\begin{eqnarray}
\widetilde\lambda^{(+)}_{(\eta_{i})}...\Gamma_{i,i+1}...
\lambda^{(-)}_{(\eta_{i+1})}=\nonumber\\
\sum_{\eta_{i},\eta_{i+1}}\widetilde\lambda^{(+)}_{(\eta_{i})}...\epsilon_{\eta_{i},\eta_{i+1}}
... \lambda^{(-)}_{(\eta_{i+1})}
\end{eqnarray}
  It corresponds to pseudoscalar meson wave functions
   $\widetilde \Psi^{(i)}\gamma_5\Psi^{(i+1)}$.
  It is worth be noted that the minimal possible mass of K-meson
is very near to the real K-meson mass.

   Further it is possible to use this structure (61) for
definition of an G-even part of nucleon emission vertices with
corresponding supercurrent conditions.
\begin{eqnarray}
&& \hat{V}^{(+)N}_{i,i+1}(z_i)\ =\ z_i^{-L_0}\{G_r, \widetilde
W^{(+)N}_{i,i+1}\}z_i^{L_0}\ ,\nonumber\\
&&\widetilde W^{(+)N}_{i,i+1}=\left[G_r,\hat W^{(+)N}_{i,i+1}\right]\\
&& \hat
W^{(+)N}_{i,i+1}=\hat{R}^{out}_{i}\hat{R}_{NS}\hat{R}^{(+)N}_{i+1,in}\nonumber
\end{eqnarray}
   Here we have i-th edging surface for usual (u,d) quark flavours
with $k_{i}^2=\mu^2$;$k_{i}k_{i+1}=0;\mu ^2\rightarrow 0$;
$\xi_{i}^2=\frac1{2}$ \\
and i+1-th edging surface for diquark quantum numbers\\
with
\begin{eqnarray}
k_{i+1}^2=k_N^2=\frac 3{2};\xi^{+(N)}_{i+1}=-\xi_i=-\sqrt{\frac
1{2}}
\end{eqnarray}
  These parameters satisfy the equation (65).

Let's note that $k_N^2=\frac 3{2}$ corresponds $m_N\approx 948Mev$
for our choice $\frac 1{\alpha'}=1,2Gev^2$.

  There are two orthogonal light-like supercurrent conditions for
G-even parts of nucleon emission vertices:
 the old one
\begin{eqnarray}
k_i\widetilde f^{(i)}= k_i (f^{(i)} + \hat{\beta}_{i} H);\nonumber\\
k_i \widetilde Y^{(i)}= k_i (Y^{(i)}+ \hat{\beta}_i \partial X )
\end{eqnarray}

 and the second one
\begin{eqnarray}
&&\widetilde f^{(N)} = k_{i+1}(f^{(i+1)} + \hat{\beta}_{i+1}H)
-\xi_i \Phi^{(i)}\nonumber\\&&+ \xi_{i+1}\Phi^{(i+1)}+(\xi_{i+1}\hat{\alpha}_{i+1}-\xi_i\hat{\alpha}_i)\Theta;\nonumber\\
&& \widetilde Y^{(N)} = -k_{i+1}(Y^{(i+1)}+ \hat{\beta}_{i+1}\partial X )- \xi_i J^{(i)}\nonumber\\
&&+\xi_{i+1} J^{(i+1)} + (\xi_{i+1}\hat
{\alpha}_{i+1}-\xi_i\hat{\alpha}_i)I
\end{eqnarray}

G-odd parts of nucleon emission vertices require third type of
vertices. Namely we take for them the following structures:
\begin{eqnarray}
&& \hat{V}^{(-)N}_{i,i+1}(z_i)\ =\ z_i^{-L_0}\left[G_r,\hat
{\widetilde
W}^{(-)N}_{i,i+1}\right]z_i^{L_0}\ ,\nonumber\\
&&\hat {\widetilde W}^{(-)N}_{i,i+1}= k_i\widetilde
f^{(i)}\widetilde f^{(N)}(\widetilde Y^{(N)})^{-\frac{1}{2}}
W^{(-)N}_{i,i+1}\nonumber\\
&&W^{(-)N}_{i,i+1}=\hat{R}^{out}_{i}\hat{R}_{NS}\hat{R}^{in}_{i+1}
\end{eqnarray}

Again we have here  previous two orthogonal light-like supercurrent
conditions :
 the first one (69): \\

    $k_i\widetilde f^{(i)}; k_i \widetilde Y^{(i)}$\\

      with $k_{i}^2=\mu^2$;$k_{i}k_{i+1}=0;\mu ^2\rightarrow
0$;$\xi_{i}^2=\frac1{2}$

  and the second one (70): \\

  $\widetilde f^{(N)}; \widetilde Y^{(N)}$ \\

        with

$k_{i+1}^2=k_N^2=\frac 3{2};\xi^{+(N)}_{i+1}=-\xi_i=-\sqrt{\frac
1{2}}$

which satisfy the equation (64).

\section{Conclusion}
    So we have simple hadron vertices in the consistent composite string model for hadron
interactions. It provides a possibility to analyse properties of
hadron amplitudes in this model.

\end{document}